\documentclass[conference,letterpaper]{IEEEtran}

\usepackage{cite}
\usepackage{amsmath,amssymb}
\usepackage{graphicx}
\usepackage{booktabs}
\usepackage{array}
\usepackage{multirow}
\usepackage{url}
\usepackage{xspace}
\usepackage{balance}
\usepackage{microtype}
\usepackage[hidelinks]{hyperref}
\usepackage{xcolor}

\newcommand{\name}[1]{\def\systemname{#1}}
\name{Diamond Agent}
\newcommand{\sys}{\systemname\xspace}

\newif\ifdraft
\drafttrue
\ifdraft
\newcommand{\zhao}[1]{{\textcolor{cyan}    { ***Zhao:      #1 }}}
\newcommand{\outline}[1]{{\textcolor{blue}    { ***Outline:      #1 }}}
\newcommand{\revision}[1]{ {\textcolor{red}    {\bf #1 }}}
\else
\newcommand{\zhao}[1]{}
\newcommand{\outline}[1]{}
\newcommand{\revision}[1]{}
\fi

\IEEEoverridecommandlockouts

\makeatletter
\def\@IEEEpubidpullup{4.0\baselineskip}
\makeatother

\begin{document}

\title{\systemname: Agentic Control of Federated HPC Resources as a Service}

\author{
\IEEEauthorblockN{
Haotian Xie\IEEEauthorrefmark{1},
Junlin Chen\IEEEauthorrefmark{1},
Mingkai Zheng\IEEEauthorrefmark{1},
Yifan Zhu\IEEEauthorrefmark{2},\\
Minu Mathew\IEEEauthorrefmark{3},
Max Burnette\IEEEauthorrefmark{3},
Yadu Babuji\IEEEauthorrefmark{4}\IEEEauthorrefmark{6},\\
Volodymyr Kindratenko\IEEEauthorrefmark{3},
Shivaram Venkataraman\IEEEauthorrefmark{5},
Kyle Chard\IEEEauthorrefmark{4}\IEEEauthorrefmark{6},
Ian Foster\IEEEauthorrefmark{4}\IEEEauthorrefmark{6},
Zhao Zhang\IEEEauthorrefmark{1}
}

\IEEEauthorblockA{
\IEEEauthorrefmark{1}Rutgers University.
Email: \{haotian.xie, junlin.chen110, mingkai.zheng, zhao.zhang\}@rutgers.edu\\
\IEEEauthorrefmark{2}University of Rochester.
Email: yifanzhu@rochester.edu\\
\IEEEauthorrefmark{3}National Center for Supercomputing Applications.
Email: \{minum, mburnet2, kindrtnk\}@illinois.edu\\
\IEEEauthorrefmark{5}University of Wisconsin--Madison.
Email: shivaram@cs.wisc.edu\\
\IEEEauthorrefmark{4}University of Chicago.
Email: \{yadunand, chard, foster\}@uchicago.edu\\
\IEEEauthorrefmark{6}Argonne National Laboratory
}
}

\IEEEpubid{%
  \makebox[\textwidth][l]{%
    \parbox[b]{\columnwidth}{%
     \vspace*{0.7\baselineskip}%
      \scriptsize
      \raggedright
      \textcopyright\ 2026 IEEE. Personal use of this material is permitted.
      Permission from IEEE must be obtained for all other uses, in any
      current or future media, including reprinting/republishing this
      material for advertising or promotional purposes, creating new
      collective works, for resale or redistribution to servers or lists,
      or reuse of any copyrighted component of this work in other works.
    }%
    \hspace{\columnsep}%
    \makebox[\columnwidth]{}%
  }%
}

\maketitle

\begin{abstract}
Efficiently aggregating and orchestrating computing power across heterogeneous clusters for HPC workflows faces four practical challenges: preserving workflow context across independently administered clusters, moving large datasets between sites, reasoning about site-specific environments and scheduler policies, and exploiting live queue and resource states for efficient task scheduling.  
To this end, we design \sys, an agentic system that enables intelligent execution of HPC workflows across heterogeneous clusters with typed skills as the interface. 
\sys provides an agent-facing workspace and skills that unify cross-site resource discovery, resource specification, data movement, task execution, and result retrieval. 
A centralized \sys instance can operate multiple supercomputers without being deployed separately on each login node. 
\sys translates high-level agent actions into valid site-specific executions, moves data through Globus Transfer, and uses live system capability and queue information to select feasible placements. 
Its event-driven continuation mechanism decouples agent actions from long-running batch jobs: persistent services monitor remote execution and resume the agent only when a result or decision-relevant event is available.
We experiment with 27 hours of telemetry and 19 matched multi-site submission rounds comprising 83 jobs across four production supercomputers. 
Compared with a fixed-site baseline,
\sys reduces the median additional completion time relative to the fastest observed placement from 42 seconds to 4 seconds, a 10.5$\times$ reduction.
\end{abstract}


\section{Introduction}
Supercomputer users are adopting coding agents, such as Codex~\cite{openai2026codex}, to program, debug, and execute workflows.
However, deploying agents across independently administered HPC systems introduces several practical challenges.
First, those with access to multiple supercomputers need to host and manage individual agents on each machine. 
Managing jobs and data across heterogeneous supercomputers remains challenging because context and execution state are fragmented across specific agent instances.
Second, running the coding agent on the login node imposes a heavy load because it requires long-running processes that communicate with the large language model service and operate the supercomputer.
Some computing centers have limited the usage of AI agents on login nodes~\cite{tacc2026policy}.



To address the above challenges, we present Diamond Agent, an agentic extension of the Diamond~\cite{xie2025diamond} service. 
The Diamond platform provides a unified interface for configuring runtime environments, submitting jobs, monitoring jobs, managing results, and synchronizing large files across supercomputers. 
It delegates execution through user-authorized Globus Compute~\cite{chard2020funcx} endpoints and uses Globus Transfer for data movement between systems~\cite{chard2017globus}. 
Diamond Agent extends this execution paradigm with an agentic system design: 
A Diamond Agent instance maintains a shared global resource view, acts through typed operations, and reasons over a single persistent task and result namespace across all authorized systems.

The agent expresses a task via a ResourceEnvelope data structure that contains the minimum viable resources, maximum useful resources, and compatibility constraints. 
The user may provide the envelope directly, or an optional AI assistant may draft it from the task description when the exact shape is uncertain. 
The draft is auditable rather than authoritative. 
Deterministic rules enforce authorization, credit, architecture, accelerator, capacity, wall-time, and site policy, while a best-feasible policy selects the concrete system and queue.

Diamond Agent uses event-driven continuation.
After receiving a durable handle, the agent can become inactive while remote execution continues.
Events such as job completion reactivate the agent when further processing is required.
A persistent Task Orchestrator owns the remote lifecycle outside the agent loop, including monitoring, bounded retry, re-placement of eligible attempts, and task-group progress. 
A result, milestone, or unresolved exception creates a durable event that reactivates the agent with compact state and provenance. 
The agent can therefore operate several supercomputers as a single logical workspace without remaining on a login node, polling Slurm, or manually reconstructing cross-site state.

We use \emph{Federated HPC Resources as a Service} to describe this user-facing control-plane abstraction. 
The underlying machines remain independent and retain authority over admission, priority, fair sharing, accounting, and execution. 
Diamond Agent unifies the agent's observations, actions, and memory within the subset of resources the user is authorized to access.
To examine the design effectiveness, we evaluate \sys using 27 hours of telemetry and live experiments on four production supercomputers.
The matched submission experiment contains 19 rounds and 83 jobs.
The experiments evaluate resource specification, live placement, adaptive multi-system execution, and end-to-end agent continuation.
Compared with a fixed-site baseline, Diamond Agent reduces the median additional completion time relative to the fastest observed placement from 42 seconds to 4 seconds.
We also evaluate the event-driven feature of \sys with two coding-agent workflows.
In this experiment, Diamond Agent was reactivated after 37 and 84 minutes of remote execution intervals with zero model-side status calls.


This paper makes four contributions:
\begin{itemize}
    \item We define an agent-facing HPC execution contract through which one agent operates multiple supercomputers using typed skills, stable task handles, unified result manifests, and shared cross-site context.
    \item We introduce event-driven continuation, which decouples active agent execution from the lifetime of a batch job and resumes the agent from durable events rather than via model-side polling.
    \item We design a user-scoped federated resource service with a Global Resource View, AI-assisted yet auditable ResourceEnvelope drafting, deterministic feasibility and credit enforcement, and the best feasible placement.
    \item We implement persistent task and group orchestration with restart-safe state, site-specific execution, bounded autonomous adaptation, and agent continuation, and evaluate the complete system across four production supercomputers.
\end{itemize}

\section{Background and Related Work}

The Globus resource-management architecture showed that autonomous centers can be federated without replacing schedulers. A broker accepts a high-level resource request, discovers available systems, and maps the request to site-specific resource managers~\cite{czajkowski1998gram}. Globus Compute provides a modern delegated execution substrate through user-authorized endpoints~\cite{chard2020funcx}. Diamond builds a common control plane above these endpoints: Image Manager prepares reusable environments, Task Manager tracks jobs and results, and Supercomputer Manager translates common operations into site-specific accounts, partitions, paths, and scheduler directives~\cite{xie2025diamond}. Diamond Agent adds live resource discovery, constraint-aware placement, and persistent cross-site adaptation.

ResourceEnvelope is related to Kubernetes resource requests and limits~\cite{kubernetes2026resources} and to SkyPilot's task-resource binding~\cite{yang2023skypilot}. Unlike Kubernetes, a Slurm job receives one fixed allocation before execution, so the envelope defines a range from minimum viable to maximum useful resources and Diamond Agent chooses one concrete request. Unlike cloud brokerage, the candidate set is limited by existing allocations, fixed node shapes, site policy, current queue state, and nonfungible per-site credits.

Parsl and Balsam provide durable workflow execution~\cite{babuji2019parsl,salim2019balsam}; queue models and resource schedulers optimize decisions within or for known systems~\cite{mao2016deeprm,li2024mrsch,brown2022wait}; and Open OnDemand, Tapis, and agentic science workbenches provide user-facing interfaces and higher-level automation~\cite{hudak2018ondemand,stubbs2021tapis,anthropic2026science}. \sys is complementary: it exposes durable cross-site execution, placement, data movement, and results through an agent-facing interface while leaving the underlying schedulers independently administered.

\section{Why an HPC Agent Is Needed}

\textbf{One scientific workflow spans several autonomous systems.}
A researcher may use one machine for CPU preprocessing, another for GPU training, and a third for evaluation or parameter sweeps. The systems expose different accounts, partitions, accelerator types, CPU-to-GPU ratios, and storage paths. A conventional coding agent attached to one login node sees only a fragment of this execution context. An HPC agent instead needs one action space and persistent workflow context over all authorized resources.

\begin{figure}[t]
    \centering
    \includegraphics[width=\columnwidth]{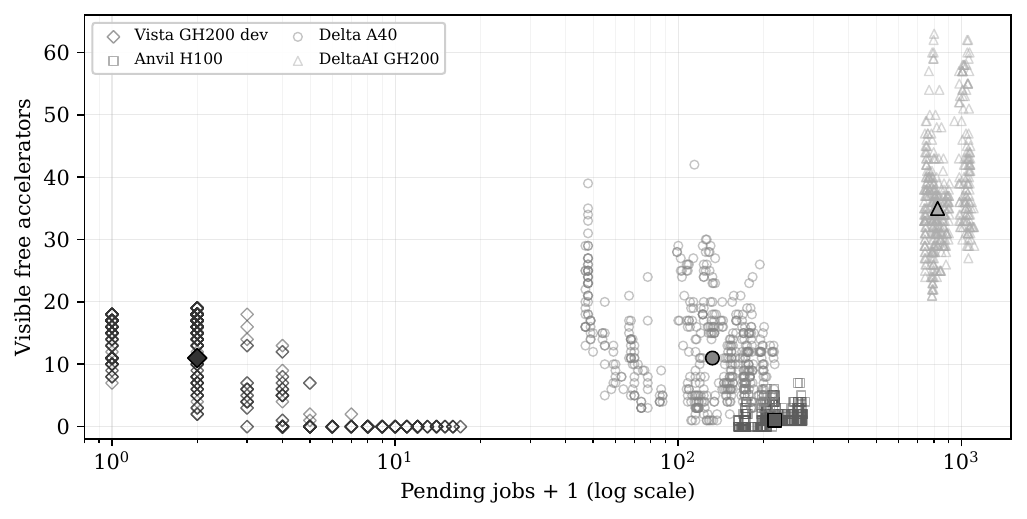}
    \caption{Pending jobs and allocatable accelerators in the four-system monitoring window. Delta, DeltaAI, and Anvil use Generic Resources(GRES) level free-device counts. Vista \texttt{gh-dev} uses idle nodes converted at one GH200 per node. Larger filled markers denote medians.}
    \label{fig:queue-state}
\end{figure}

\textbf{Resource intent and live capacity must be reconciled.}
To avoid batch jobs failing after a long waiting time, users tend to request more resources than necessary.
Yet a larger request has fewer feasible placements and fewer backfill opportunities~\cite{jette2023slurm}. 
Queue length alone is also insufficient. In our four-system trace, DeltaAI's GH200 partition had a median of 823 pending jobs while exposing a median of 35 free devices; Anvil's H100 partition had fewer pending jobs but only one free device; Vista's GH200 development partition had a median of one pending job and 11 available devices. Figure~\ref{fig:queue-state} shows that an agent must combine the task's resource bounds with queue state and allocatable capacity rather than follow a fixed site preference.

\textbf{Batch execution can outlive active agent execution, and the execution path can change.}
Supercomputer center notices show that compute, authentication, storage, and access services may change independently~\cite{nersc2026status,nersc2026timeline,tacc2026updates}. More generally, a queued task can remain pending while another authorized system becomes a better placement. The agent should not stay active merely to watch this state. It needs a persistent control loop that can observe remote jobs, preserve their context, apply bounded policies, and return control only when the workflow may need to move forward.
These observations motivate four requirements.
\textbf{R1: cross-site workflow continuity.}
A single agent must preserve workflow context and data continuity while acting across multiple authorized supercomputers. Task state, results, and artifacts produced on one system must remain accessible to later workflow stages on another, without maintaining a separate agent context at each site.
\textbf{R2: asynchronous agent execution.}
Long-running HPC execution must be decoupled from active agent execution. After submitting remote work, a persistent service must maintain its state and trigger the agent only when a new event requires its attention.
\textbf{R3: heterogeneous system abstraction.}
The agent must be able to express HPC actions through a common interface, while the system translates them into site-specific accounts, queues, resource requests, execution environments, and scheduler directives.
\textbf{R4: live-state-aware placement.}
The system must expose current resource capabilities and queue conditions across the federation and use them to map each task to an efficient, policy-compliant execution site.

\section{Agentic Interaction Model}

Diamond Agent treats an HPC workflow as a sequence of agent actions separated by asynchronous remote actions. The agent follows an observe--plan--act--yield--resume cycle. 
It observes prior task state, plans an action, invokes a typed skill, yields while the control plane executes the action, and resumes when a new response changes the next decision. 
Placement is deterministic; agentic behavior lies in maintaining this cross-site action and reasoning loop without keeping the agent active during remote execution.

\subsection{One Agent, Multiple Supercomputers}

The agent runs on the user's own machine or a normal agent host, not separately on every login node. It sees site-independent identifiers for tasks, groups, attempts, events, and results. Site-specific accounts, partitions, QoS fields, scheduler commands, endpoint identifiers, and file paths remain available in the audit trace but are translated by Diamond rather than remembered by the agent. This lets the agent use Delta for one stage, Vista for another, and Anvil or DeltaAI for sibling tasks while retaining one scientific objective and one result history.

Table~\ref{tab:agent-ops} summarizes the core skill contract. The same typed operations can be exposed as agent skills, tool calls, or ordinary REST calls. Submission returns immediately with a durable handle and placement trace. Result access returns a compact manifest with provenance and locations; large logs and artifacts are fetched only when the agent needs them.

\begin{table*}[t]
\caption{Agent-facing operations and asynchronous continuation semantics.}
\label{tab:agent-ops}
\centering
\scriptsize
\begin{tabular}{@{}p{0.17\textwidth}p{0.24\textwidth}p{0.27\textwidth}p{0.24\textwidth}@{}}
\toprule
\textbf{Operation} & \textbf{Agent intent} & \textbf{Immediate response} & \textbf{Later trigger condition} \\
\midrule
\texttt{inspect\_federation} & Observe systems, queues, capacity, health, and credit scope & Timestamped normalized snapshot & None \\
\texttt{draft\_envelope} & Convert a task description into auditable resource bounds and constraints & Typed draft, rationale, confidence, validation result & None \\
\texttt{submit\_task} & Execute one scientific or engineering action & Task handle, selected placement, granted shape, decision trace & Completion, failure, unresolved policy event, or requested milestone \\
\texttt{submit\_group} & Fan out related tasks while preserving one objective & Group and child handles, placements, round state & Round terminal, stalled child, or group exception \\
\texttt{move\_data} & Stage or replicate large inputs and outputs across sites & Transfer handle, source and destination references, policy trace & Transfer completion, failure, or requested milestone \\
\texttt{get\_result} & Perceive outputs and provenance & Result manifest, attempt history, selected metadata & None \\
\texttt{replan}/\texttt{cancel} & Adapt the operational plan & Updated attempt or terminal state & Next terminal or exception event \\
\bottomrule
\end{tabular}
\end{table*}

\subsection{Event-Driven Continuation}

We use an \emph{event-driven continuation} policy to decouple active agent execution from long-running HPC jobs. After \texttt{submit\_task} or \texttt{submit\_group}, the agent receives a durable handle and can release active execution while the control service continues monitoring the remote work at its configured rate without agent calls. When a registered condition occurs, the Task Orchestrator appends an event with a stable event identifier. A runtime adapter or connector deduplicates the event and reactivates the agent with a compact payload containing the task or group handle, new state, selected system, attempt summary, result manifest, and allowed next actions.
Event-driven continuation does not require the agent to remain active while remote work is executing. The agent process may suspend, exit, or restart, while task state remains persistently managed by the control service. Event delivery follows at-least-once semantics, and stable event identifiers allow the runtime adapter to suppress duplicate reactivations. The system does not guarantee exactly-once external side effects; subsequent actions therefore use stable task identifiers and idempotency checks. Reactivation latency is determined by the control-service interval and connector delay rather than by continuous agent-side polling.

\subsection{Persistent Context and Bounded Autonomy}

The persistent store serves as operational memory outside the model context window. A resumed task can inspect the original objective, ResourceEnvelope, placement decision, site provenance, sibling states, prior attempts, and result locations without replaying the entire conversation. Task groups preserve relationships among CPU preprocessing, GPU training, evaluation, or parameter-sweep children even when they run on different sites. This shared context is what makes the federation agentic rather than a set of unrelated remote submissions.

Diamond Agent also separates workflows from infrastructure policy. The agent may choose the next experiment, interpret results, change parameters, or request a replan. Deterministic rules enforce authorization, site policy, credit, resource compatibility, retry budgets, and blocked operations. Routine recovery can proceed without invoking the model. Events reactivate the agent only when an unresolved exception occurs or a requested job result becomes available. Credentials remain in the authorization path and are never inserted into the model prompt.

\section{Scheduling Formulation}

\subsection{ResourceEnvelope and Candidate Set}

A task $t$ is described by a ResourceEnvelope $E_t$ containing GPU, CPU, memory, node, and wall-time ranges together with architecture, accelerator, container, data, and policy constraints. 
For numerical specifications (e.g., GPU and CPU), a ResourceEnvelope requires the minimum and maximum values.
The minimum is the lower bound of resources required to run correctly; the maximum is the upper bound on resources expected to provide useful speedup. An envelope may be user-authored or AI-drafted. An AI draft also stores its source, confidence, and rationale, but these fields do not bypass schema validation or deterministic rules. The current implementation uses independent ranges and then selects a single fixed batch allocation before submission.

A placement candidate is

\begin{equation}
  c=(s,p,a,z),
\end{equation}

where $s$ is a system, $p$ a partition, $a$ an account, and $z$ a concrete job size selected from the envelope. Let $X$ be the current ResourceSnapshot. Candidate $c$ is feasible when all deterministic rules pass and the predicted charge fits the user's site-specific budget:

\begin{equation}
  \mathcal{F}_t=\{c \mid R_j(E_t,c,X)=1\ \forall j,\ \widehat K(t,c)\le B_s(t)\}.
\end{equation}

The rules cover endpoint health, authorization, blocked partitions, processor architecture, accelerator type and memory, minimum capacity, wall time, container compatibility, and user limits. Credits are not treated as one fungible global currency. $B_s(t)$ is the amount of the allocation on site $s$ that the user permits this task to consume.

\subsection{Best-Feasible Placement}

For the default earliest-completion objective, Diamond Agent computes

\begin{equation}
  \widehat C(t,c)=\widehat W(t,c)+\widehat R(t,c)+\widehat E(t,c),
\end{equation}

where $\widehat W$ is a deterministic queue-delay estimate, $\widehat R$ is a declared or historical runtime estimate for the granted shape, and $\widehat E$ is the measured preparation cost when the required environment is absent. In live mode, the wait estimator first uses request-sized slots visible in free GRES or CPU capacity and then queue depth; it does not infer free GPUs from node state alone. 
Ties are resolved by a lower fraction of the site's remaining allowed credit, lower specialized-resource waste, and finally, user preference. The selected candidate is

\begin{equation}
  c^*=\arg\min_{c\in\mathcal{F}_t}\widehat C(t,c).
\end{equation}

For a fixed snapshot and feasible candidate set, the policy enumerates all candidates and chooses the one with minimum estimated completion time. This does not imply minimum actual completion time or global optimality across interacting or future tasks.

\section{System Design}

\begin{figure*}[t]
    \centering
    \includegraphics[width=0.96\textwidth]{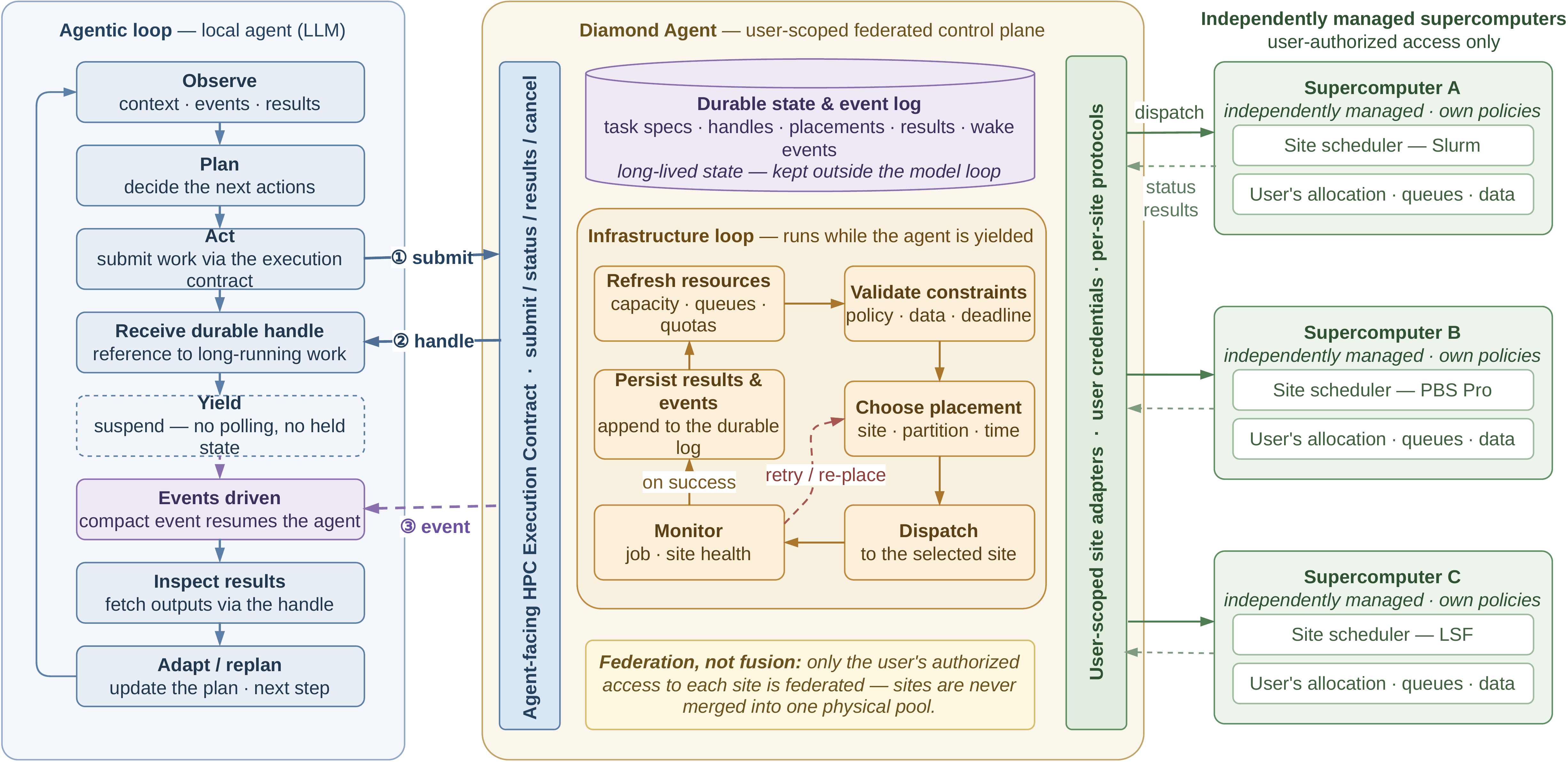}
    \caption{Diamond Agent Architecture}
    \label{fig:architecture}
\end{figure*}

Figure~\ref{fig:architecture} separates the system into an agent-facing control loop and an infrastructure control loop. The Agent Gateway exposes typed skills and stable identifiers. The Federated Resource Service maintains current state and chooses a valid placement. The persistent Task Orchestrator owns asynchronous execution and creates events. Diamond and Globus Compute provide delegated, site-specific actuation. This separation lets the agent reason over one logical workspace while deterministic services handle long-lived infrastructure state.

\subsection{Agent Gateway and Unified Context}

The Agent Gateway maps agent actions into typed operations and returns task logs. It stores the objective, envelope, decision trace, attempt history, event cursor, and result references outside the model context. Runtime-specific adapters may package the operations as reusable skills, but the operation schema is independent of Claude Code, Codex, or any other agent runtime. The gateway also bounds response size: task submission returns control metadata, while logs and large artifacts remain referenced by the result manifest until explicitly requested.

\subsection{Global Resource View and Placement}

The Global Resource View contains one record per authorized system, partition, and account. Static fields describe architecture, accelerator type and memory, node configuration, wall-time caps, container support, and site policy. Dynamic fields describe endpoint health, node-state buckets, GRES-level free accelerators, free CPUs, global and user queue counts, and timestamps. A degraded endpoint remains visible with health state and known metadata, but stale or offline records cannot pass placement checks.

A ResourceEnvelope may be supplied by the user, inherited by a task group, or drafted by the agent from task text and code context. The draft is a typed proposal with a rationale and confidence value, not a placement decision. The scheduler itself is model-free. Deterministic rules construct the feasible set, resource sizing chooses one legal request between the minimum viable and maximum useful bounds, and the declared objective selects the best candidate. Every envelope source, granted size, rejection reason, estimate, credit check, and tie-break decision is recorded for later agent inspection.

\subsection{Data Movement and Artifact References}

Diamond Agent inherits Diamond's native Globus Transfer path for large-file movement~\cite{xie2025diamond,chard2017globus}. Inputs and outputs are represented by logical data references that resolve to authorized Globus collections and paths. When the selected system does not already hold an input, Diamond can stage the data before dispatch; completed outputs can be returned to a designated collection or replicated to the system used by the next task. The underlying transfer remains a direct managed transfer between storage endpoints rather than a file stream through the agent.

Data movement follows the same agent-facing lifecycle as compute. A transfer returns a durable handle, its state is monitored outside the model loop, and completion or failure can produce a new event. Task and group records retain transfer identifiers, source and destination references, and artifact provenance. Result manifests expose locations and metadata instead of inserting large datasets, checkpoints, or images into the model context. Thus one agent can coordinate both computation and data dependencies across sites without manually copying files or maintaining a separate transfer session.

\subsection{Event-Driven Orchestration and Task Groups}

The Task Orchestrator turns a placement into a durable task handle and invokes Diamond's execution layer. It records scheduler identifiers, endpoint bindings, state transitions, retry budgets, continuation registrations, and events. Pending or failed work may be retried or re-placed under explicit policies, while running jobs remain bound to their original scheduler allocation. Multiple attempts are tracked separately so that the resumed agent sees one coherent task history rather than site-specific fragments.

A parent task may create independently placed children that share an objective, envelope, round, and policy. The orchestrator aggregates child states and results, may relax a harmful soft placement constraint, and starts later rounds only after the required children become terminal. The agent receives one group-level response rather than polling each scheduler or keeping one agent on each site. Routine state changes remain inside the orchestrator; registered milestones and exceptions become new events.

\section{Implementation}

\subsection{Agent Gateway and Event Bridge}

The service exposes federation, envelope, task, group, result, replan, cancel, and event interfaces. Runtime adapters convert these operations into skills or tool calls without changing the backend contract. Submission is synchronous only through placement and remote dispatch; it returns a handle as soon as the scheduler accepts the job. The agent can then become inactive. Each task stores a continuation policy and an event cursor. An event bridge reads durable events, suppresses already-consumed event identifiers, and supplies the next agent action with a compact continuation payload rather than the full task log.

The payload contains identifiers, the new state, placement and attempt provenance, a result manifest, and a bounded set of next actions. Large stdout, checkpoints, images, and scientific artifacts are not injected into the prompt automatically. The agent requests only the records needed for interpretation. This keeps cross-site context available while controlling token growth.

\subsection{Persistent Control Loop}

Diamond Agent runs as one control service. Each control cycle first refreshes federation state and then advances tasks and groups against the resulting snapshot. Remote actions are delegated through Globus Compute, allowing the agent to operate several sites through one control loop while each center retains its own scheduler and access boundary.

Tasks, groups, scheduler identifiers, retry counters, continuation registrations, consumed event identifiers, placement traces, and timelines are persisted. At startup, the service restores these records and re-binds pending and running jobs to their original endpoints. The resource cache is rebuilt on the first refresh, while scheduler accounting remains authoritative for final state and timing. The agent runtime, control service, and remote batch jobs can therefore have independent lifetimes.

\subsection{Globus Transfer Integration}

Diamond's data manager submits stage-in, stage-out, and site-to-site synchronization requests through Globus Transfer. Each request records the source collection and path, destination collection and path, transfer task identifier, and relationship to the compute task. A compute attempt can depend on successful stage-in, and its result manifest can register stage-out or replication targets for later tasks. Transfer progress is therefore part of the persistent task timeline and can survive an agent or service restart.

The agent is not in the data path. It issues a typed transfer intent and later receives a compact transfer status or artifact reference. This design is important for large files: movement proceeds through Globus-managed storage endpoints, while the model handles only metadata and decisions. The current placement objective does not yet jointly optimize transfer duration with queue wait and runtime; it executes the requested movement reliably after placement.

\subsection{Federated Resource Collection}

Each site contributes a static profile and a live probe. The profile records hardware specifications and executable policy, including partitions, accounts, QoS requirements, node shapes, accelerators, architecture, wall-time caps, container support, and scheduler-dialect switches. The live probe collects partition node states, free CPUs, free accelerators, global and user queue counts, available accounts, architecture, and endpoint health through the user-authorized endpoint. 

Free GPU capacity is computed from configured and used GRES on usable nodes. This is important on shared GPU nodes, where node state alone can hide or overstate allocatable devices. The Federation Collector merges live values with the profile into a timestamped ResourceSnapshot and preserves an explicit availability state for every known execution path. A bounded history supports monitoring, estimator input, and retrospective analysis.

Beyond the four systems used in the evaluation, the current site-profile interface is also configured for NERSC Perlmutter and ALCF Sophia and Polaris.

\subsection{Constraint-Aware Placement}

The optional envelope generator asks the agent for a structured draft rather than free-form prose. The response contains numeric lower and upper bounds, compatibility constraints, objective, rationale, and confidence. Schema validation checks the ranges, and the task record preserves both the original draft and the normalized envelope. The generator is an assistance path for uncertain users; the scheduler itself is deterministic and can operate entirely from a user-authored envelope.

Submission then validates the ResourceEnvelope and evaluates nine deterministic rule categories for every system-partition record. Hard failures remove a candidate before sizing or objective evaluation. Authorization, blocked partitions, architecture, accelerator and memory requirements, capacity, wall time, container support, user limits, and site-specific credit are enforced in this stage.

For each feasible record, sizing starts from the maximum useful request, reduces it to observed and site-allowed capacity when needed, and never falls below the minimum viable request. The live wait estimator uses free GRES or CPU capacity before node-idle state, converts capacity into request-sized slots, and combines this with queue pressure. Runtime comes from the task declaration or previous runs; environment preparation is included only when the service has an explicit cache-state cost. The policy enumerates all candidates and selects the minimum estimated completion time, with deterministic credit, waste, and preference tie breakers. The decision trace is stored so the agent can explain or revise the plan later.

\subsection{Task State Machine and Autonomous Adaptation}

The selected request is translated to the site's scheduler dialect and submitted through the chosen endpoint. A successful submission creates a persistent job handle and advances through submitted, pending, running, and terminal states. Each control observes endpoint and scheduler state and evaluates continuation policies without invoking the model.

Bounded policies may retry failed work, enlarge wall time, choose another valid resource shape, or re-place a pending attempt when another feasible placement is strictly better. Attempts remain distinct and are reconciled against the scheduler state before one result is accepted as current. Manual agent re-planning uses the same path and can exclude selected systems. Only registered milestones, unresolved exceptions, and requested results are emitted as events. This division lets the service adapt to infrastructure changes while reserving agent calls for scientific decisions.

\subsection{Remote Execution, Persistence, and Safety}

The execution layer resolves the account, partition, QoS, memory and accelerator directives, working directory, and environment command from the site profile. 
It submits the generated batch request, parses the scheduler identifier, and uses the bound endpoint for status and cancellation. Final wait and runtime are audited from scheduler accounting rather than inferred from the polling interval.

The persistent store contains task and group records plus an append-only event stream. A separate experiment log records resource refreshes, envelopes, decisions, dispatches, continuation events, and scheduler audits. Cluster credentials, Globus tokens, SSH keys, and project secrets remain in the authorization and execution path. The agent receives authorized operation results, not reusable credentials. 
The implemented submission path enforces authorization, blocked-partition rules, compatibility checks, wall-time and resource ceilings, site credit, and retry budgets, but does not sandbox user code.

\section{Experiment and Evaluation}

Our evaluation examines whether \sys provides a practical and effective execution substrate for agent-driven workflows across federated HPC systems: maintaining a usable cross-site resource view, translating uncertain resource intent into executable requests, selecting effective placements from live system state, adapting multi-system execution, and continuing agent workflows asynchronously over long-running HPC jobs. 
We seek to answer the following five questions with experiments:

\begin{itemize}
    \item \textbf{RQ1:} Can the service maintain a useful multi-system resource view with acceptable overhead and failure visibility?
    \item \textbf{RQ2:} Can AI draft executable ResourceEnvelopes, and how do their bounds behave under real scaling and queue conditions?
    \item \textbf{RQ3:} How close is \sys's live-state placement to the fastest observed placement in matched multi-site submissions?
    \item \textbf{RQ4:} Can the service execute and adapt a live multi-supercomputer task group and recover from endpoint failure?
    \item \textbf{RQ5:} Can \sys remain inactive during long HPC actions and resume from results without agent-side status polling?
\end{itemize}

\subsection{Testbed and Method}

Table~\ref{tab:testbed} summarizes the experiment platforms. It spans four systems at three centers, nine partitions, x86\_64 and Arm architectures, and five accelerator configurations. All queue probes and live jobs used the same Diamond and Globus Compute path as normal operation. The long monitoring run covers 27 hours.

\begin{table}[t]
\caption{Federation used in the evaluation.}
\label{tab:testbed}
\centering
\scriptsize
\setlength{\tabcolsep}{2pt}
\begin{tabular}{@{}p{0.21\columnwidth}p{0.13\columnwidth}p{0.27\columnwidth}p{0.29\columnwidth}@{}}
\toprule
\textbf{System} & \textbf{Arch.} & \textbf{Accelerators} & \textbf{Partitions / access} \\
\midrule
Delta@NCSA & x86\_64 & A100 40G, A40 48G & 3, multi-user endpoint \\
DeltaAI@NCSA & aarch64 & GH200 120G, 4/node & 1, login-node endpoint \\
Anvil@RCAC & x86\_64 & A100 40G, H100 80G & 2, login-node endpoint \\
Vista@TACC & aarch64 & GH200 96G, 1/node & 3, login-node endpoint \\
\bottomrule
\end{tabular}
\end{table}

For RQ2, the agent generated three ResourceEnvelopes from workload descriptions without manual edits. We fixed each workload to one partition to isolate sizing from placement, mechanically derived minimum, scheduler-selected, and maximum requests from the envelope, and ran three repetitions per request. Six additional scaling probes produced 33 completed Slurm jobs in total. The workloads were CPU preprocessing on Delta, single-GPU inference on DeltaAI, and single-node DDP training on DeltaAI. All requests used the same 30-minute wall time. Scheduler accounting supplies wait, runtime, and completion; in-band instrumentation supplies CPU/GPU utilization and memory use. To validate the AI's maximum bound, a resource increase is called useful when the per-doubling speedup is at least 1.25 and marginal parallel efficiency is at least 50\%.

For RQ3, at each time window, we captured one federation snapshot and submitted the same fixed-shape job to every raced candidate within 2.5 to 11.8 seconds. The campaign contains 19 completed rounds over about four hours: seven CPU rounds, six one-GPU rounds, and six four-GPU rounds, totaling 83 submissions. CPU and one-GPU jobs slept for 180 seconds; four-GPU jobs slept for 300 seconds. Slurm \texttt{sacct} provides Submit, Eligible, Start, End, and elapsed time. For each round, the fastest completed copy is the fastest observed.

\subsection{RQ1: Federated Resource View}

The collector produced 543 ResourceSnapshots and 4,737 partition records over 27 hours. Refresh duration was 64.1 seconds on average, 65.6 seconds at the median, 87.7 seconds at p95, and 288.0 seconds at the maximum. Mean data freshness was 183.3 seconds, p95 was 209.1 seconds, and the maximum was 409.2 seconds. Each round used 28.6 Globus Compute requests on average and produced about 5.0 KB of state. Over the full run, the process consumed 416 CPU seconds and reached 108.1 MB peak RSS (Resident Set Size).

These measurements show that one agent can maintain a minute-scale observation space over four independently administered systems with modest overhead. Every known execution path remains represented by identity, timestamp, and availability state, so the scheduler can exclude an unavailable path without losing the surrounding task context. The current collector is effective for meta-scheduling decisions that evolve over minutes rather than sub-second cluster scheduling.

\subsection{RQ2: AI-Assisted ResourceEnvelope Sizing}

All three drafts were produced by the AI path, with recorded confidence values of 0.86, 0.88, and 0.86. They correctly separated a CPU-only workload, a one-GPU inference workload, and a one-to-four-GPU DDP workload, and every derived request passed the production rule chain. Table~\ref{tab:ai-sizing} summarizes the live outcomes. Completion is measured from \texttt{sacct}; the selected request is the concrete shape chosen from the live snapshot, not another model output.

\begin{table*}[t]
\caption{AI-generated ResourceEnvelopes and actual sizing outcomes. Values are means over three repetitions unless noted. The 128-CPU and 2-GPU rows are two-repeat scaling probes.}
\label{tab:ai-sizing}
\centering
\scriptsize
\setlength{\tabcolsep}{3pt}
\begin{tabular}{@{}p{0.13\textwidth}p{0.23\textwidth}p{0.15\textwidth}p{0.20\textwidth}p{0.23\textwidth}@{}}
\toprule
\textbf{Workload} & \textbf{AI envelope} & \textbf{Selected shape} & \textbf{Actual completion} & \textbf{Observed implication} \\
\midrule
CPU preparation & 4--64 CPUs \newline 2--8 GB \newline no GPU & 64 CPUs \newline 2 GB & 42 s; minimum 4-CPU shape: 226 s; 128-CPU probe: 2029 s & 64 CPUs reduced runtime without a queue penalty. 128 CPUs still scaled in-kernel but waited 2012 s, so the AI upper bound was conservative for scaling and appropriate for completion. \\
Single-GPU inference & 1 GPU \newline 2--16 CPUs \newline 16--64 GB & 1 GPU \newline 16 CPUs \newline 16 GB & 146 s; minimum 2-CPU shape: 148 s & GPU utilization and runtime were unchanged across the tested CPU range, so the upper CPU bound provided no completion benefit and should be refined by later profiling. \\
DDP training & 1--4 GPUs \newline 4--32 CPUs \newline 16--128 GB & 4 GPUs \newline 32 CPUs \newline 16 GB & 345 s; 1 GPU: 58 s; 2-GPU probe: 55 s & The 1$\rightarrow$2$\rightarrow$4 GPU speedups were 1.90 and 1.96 per doubling, so 4 GPUs passed the usefulness threshold, but its longer queue wait made 1--2 GPUs better for end-to-end completion. \\
\bottomrule
\end{tabular}
\end{table*}

The experiment result in Table~\ref {tab:ai-sizing} shows why AI-generated ResourceEnvelopes should remain auditable proposals rather than final allocation decisions. The AI captured the workload classes and useful GPU scaling range, and its CPU-preparation bound led the scheduler to the best measured completion point among the main shapes. The GPU-bound workload also shows that an initially reasonable upper bound may provide no completion benefit. The DDP result further shows that a runtime-useful maximum is not necessarily completion-optimal under a batch queue. Diamond Agent therefore treats the AI envelope as an auditable resource hypothesis: live capacity, queue state, deterministic rules, and accumulated profiling select the granted shape and refine future envelopes.

As a separate constraint check, an intentionally impossible request for at least 200 GB of memory per accelerator was rejected at every site and entered an UNSCHEDULABLE state with per-candidate explanations. Thus the AI can assist resource expression, while deterministic enforcement remains the safety boundary.

\subsection{RQ3: Matched Multi-Site Placement}

For each round, completion is $C=\text{End}-\text{Submit}$, queue wait is
$W=\text{Start}-\text{Eligible}$, and additional completion time is the
selected completion minus the fastest completed copy in the same round.
Table~\ref{tab:policies} compares three deployable placement policies with the fastest observed placement. \emph{Fixed Delta} always selects Delta when it is feasible. \emph{Shortest feasible queue} selects the feasible candidate with the shortest queue in the pre-submission snapshot. \emph{Best-feasible}, the production policy, evaluates all feasible candidates and selects the one with the minimum estimated completion time using queue delay, runtime, and environment-preparation estimates. The \emph{fastest observed placement} is a post-hoc reference defined by the candidate with the shortest measured completion time in that round. All three deployable policies select only from the same feasible candidate set, and every selected candidate has a measured outcome.

\begin{table*}[t]
\caption{Actual matched-submission outcomes over 19 rounds and 83 jobs. Completion and additional completion time are measured from scheduler accounting, not model replay.}
\label{tab:policies}
\centering
\scriptsize
\begin{tabular}{@{}lrrrrrr@{}}
\toprule
\textbf{Policy} & \textbf{Fastest selected} & \textbf{Median $C$} & \textbf{Mean $C$} & \textbf{Mean wait} & \textbf{Median additional completion time} & \textbf{P95 additional completion time} \\
\midrule
Fixed Delta & 0\% & 227s & 1422s & 1198s & 42s & 5873s \\
Shortest feasible queue & 47\% & 322s & 738s & 513s & 4s & 3000s \\
Best-feasible & 47\% & \textbf{196s} & 817s & 592s & 4s & 3000s \\
Fastest observed placement & 100\% & 194s & 281s & 57s & 0s & 0s \\
\bottomrule
\end{tabular}
\end{table*}

The fixed-system has a mean additional completion time of 1,141 seconds, about 19 minutes per task, and a p95 additional completion time of 5,873 seconds. Both live-state policies reduce the typical additional completion time to four seconds and select the actual fastest candidate in 47\% of rounds. Best-feasible reaches a median completion of 196 seconds, close to the 194 seconds median of the fastest observed placements and 39\% lower than shortest queue's 322 seconds.

The tail cases remain large. Best-feasible has a mean additional completion time of 536 seconds and a p95 additional completion time of 3,000 seconds. The best-feasible policy does not dominate the shortest-queue policy: although its median completion time is lower, a few four-GPU placements waited much longer and raised its mean completion time above that of shortest queue. The matched submissions therefore support two narrower conclusions. First, live federation state can avoid the large penalty of always using one system. Second, the current estimator often identifies a near-best placement but is not yet accurate enough for larger resource requests. 
This policy minimizes the estimated objective over the feasible candidates, but does not guarantee minimum actual completion time. This experiment measures the estimation error against actual queue outcomes.

\subsection{RQ4: Live Multi-Supercomputer Execution}

\begin{figure*}[t]
    \centering
    \includegraphics[width=0.88\textwidth]{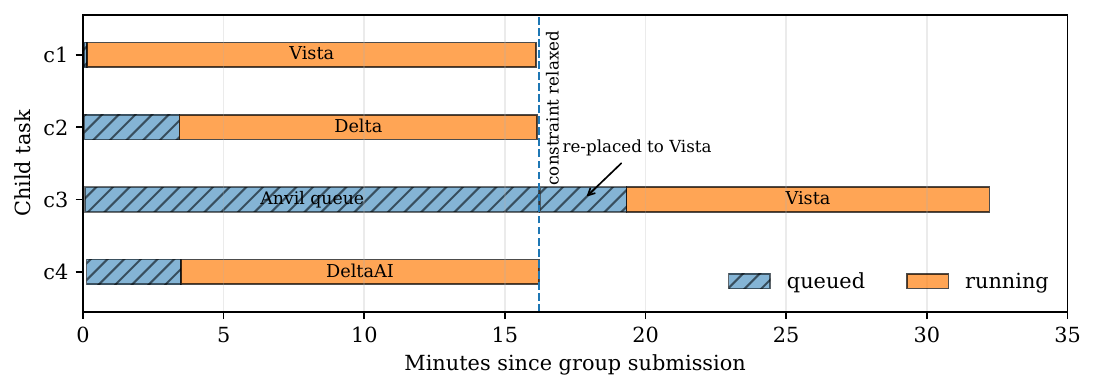}
    \caption{Live task-group adaptation. Child c3 was queued on Anvil. After its siblings completed, the soft distinct-system constraint was released and c3 was re-placed on Vista.}
    \label{fig:fanout}
\end{figure*}

We construct a SWE-RL-style scenario with four independent 15-minute single-GPU tasks representing parallel inference workers. Running these workers across different systems also exposes them to heterogeneous execution environments, which helps avoid coupling the workflow to a single site configuration. Through the agent-facing interface, the agent creates a task group with a \emph{best-effort placement diversity} goal, asking \sys to spread the children across distinct supercomputers when possible. \sys initially places the four children on Vista, Delta, DeltaAI, and Anvil. Figure~\ref{fig:fanout} shows the execution. The tasks on Vista, Delta, and DeltaAI made progress and completed, while the task assigned to Anvil remained pending. Once preserving the distinct-system goal no longer provided useful parallelism, \sys relaxed the soft constraint and re-placed the remaining task on Vista. All four tasks completed without user intervention, with final-job queue waits of at most 0.7 minutes and a group makespan of approximately 32 minutes.

We also evaluated controlled endpoint loss. The collector marked the affected execution path unavailable, the orchestrator excluded it from new placements, and a replacement was submitted to Vista 34 seconds after detection. After the endpoint recovered, Diamond Agent reconciled and cancelled the superseded attempt. The task completed without user intervention. Together, these scenarios validate fan-out, persistent group state, policy-driven re-placement, and cross-layer recovery, but not large-scale throughput.

\subsection{RQ5: End-to-End Agent Continuation}

We evaluate the agent-facing path with two running coding agents. Claude Code with Claude Opus 4.8~\cite{anthropic2026claudecode,anthropic2026opus48} used the common skill interface to deploy and fine-tune SAM 3~\cite{carion2025sam3} on one A40 GPU at NCSA Delta. Codex with GPT-5.5~\cite{openai2026codex,openai2026gpt55} used the same operation schema on one GH200 at NCSA DeltaAI. Each workflow inspected the target resources, prepared code and the batch request, submitted through Diamond Agent, became inactive after submission, resumed from a completion event and result manifest, and inspected the outputs.

\begin{table}[t]
\caption{Event-driven behavior in two end-to-end coding-agent traces. The polling column is a derived five-minute counterfactual, not a measured baseline.}
\label{tab:agent-traces}
\centering
\scriptsize
\setlength{\tabcolsep}{3pt}
\begin{tabular}{@{}p{0.20\columnwidth}p{0.25\columnwidth}rrr@{}}
\toprule
\textbf{Agent} & \textbf{Remote job} & \textbf{Async interval} & \textbf{Status calls} & \textbf{5-min polls} \\
\midrule
Claude Code & Delta, A40 & 37 min & 0 & 8 \\
Codex & DeltaAI, GH200 & 84 min & 0 & 17 \\
\bottomrule
\end{tabular}
\end{table}

Table~\ref{tab:agent-traces} shows that neither workflow made an agent call to ask whether Slurm had finished during the asynchronous interval. The control service monitored the jobs and supplied the terminal state and result manifest to the reactivated agent. A model-side polling loop at five-minute intervals would require 8 and 17 model-side status calls over the same measured intervals. This comparison is analytical and does not claim that every system without Diamond Agent must use model polling; a separate workflow engine or shell watcher could also monitor the jobs. The contribution is that monitoring, cross-site task memory, result delivery, and agent resumption are one reusable interface.

\section{Limitations and Future Work}

The current \sys implementation intentionally separates resource sizing from placement. Resource bounds are first converted into a feasible resource shape, after which the placement service selects a system and queue using live federation state and estimated completion cost. This design is simple and deterministic, but a resource shape that is computationally efficient may not be optimal once queue delay, site choice, and data movement are considered jointly. In addition, the current placement model relies on hand-designed estimators rather than continuously learning from previous executions. Finally, the implemented policy gates constrain authorization and resource consumption, but do not provide isolation for user-supplied code, which executes with the user's HPC permissions.

Our future work focuses on extending the agent-facing resource service in three directions. First, we plan to introduce online-learning-based scheduling that continuously updates queue-wait and runtime estimates from observed executions and adapts placement decisions to changing system behavior. Second, we will jointly optimize resource sizing and placement, including resource shape, queue delay, execution time, site choice, and data-movement cost. Third, we will extend the current wall-time and resource limits with workflow-level auditing and risk controls that account for model tokens, node-hours, accelerator-hours, and cumulative resource consumption. These mechanisms will allow long-running agents to operate under explicit computational and monetary budgets.

\section{Conclusion}

Diamond Agent is an agent system that enables HPC workflow execution across heterogeneous supercomputers through a unified agent-facing interface. 
It combines cross-site workflow context, data movement, heterogeneous system abstraction, live resource and queue states, and persistent execution, enabling high-level agent actions to be translated into valid operations on independently administered HPC systems.
A key design principle is to decouple active agent execution from the lifetime of remote batch jobs. 
Through event-driven continuation, compute and data-movement tasks proceed under persistent system control while the agent remains inactive, and the agent is reactivated only when results or decision-relevant events require further reasoning. 
This allows one agent to coordinate multiple HPC facilities without maintaining separate site-specific contexts or continuously polling remote jobs.
Our evaluation on production supercomputers demonstrates the practical effectiveness of this design for cross-site resource observation, resource specification, adaptive job placement, multi-system execution, and long-running agent workflows. 
Together, these results show that \sys enables agents to coordinate federated HPC resources while preserving the policies and administrative boundaries of the underlying systems.

\section*{Acknowledgments}
This research used both the DeltaAI advanced computing and data resource, which is supported by the National Science Foundation (award OAC 2320345) and the State of Illinois, and the Delta advanced computing and data resource which is supported by the National Science Foundation (award OAC 2005572) and the State of Illinois.

The authors acknowledge the Texas Advanced Computing Center (TACC) at The University of Texas at Austin for providing computational resources that have contributed to the research results reported within this paper. URL: http://www.tacc.utexas.edu

This research is supported by the OAC-2311767, OAC-2311768, OAC-2311769, and OAC-2401245.

\balance
\bibliographystyle{IEEEtran}
\bibliography{references}

@inproceedings{xie2025diamond,
  author={Xie, Haotian and Marwaha, Rohan and Mathew, Minu and Bian, Song and Yang, Gengcong and Yan, Minghao and Babuji, Yadu and Price, Owen and Wang, Yinzhi and Kindratenko, Volodymyr and Venkataraman, Shivaram and Chard, Kyle and Foster, Ian T. and Zhang, Zhao},
  title={Diamond: Harnessing {GPU} Resources for Scientific Deep Learning},
  booktitle={2025 IEEE International Conference on eScience},
  pages={196--204},
  year={2025},
  doi={10.1109/eScience65000.2025.00031}
}

@inproceedings{chard2020funcx,
  author={Chard, Ryan and Babuji, Yadu and Li, Zhuozhao and Skluzacek, Tyler and Woodard, Anna and Blaiszik, Ben and Foster, Ian and Chard, Kyle},
  title={{funcX}: A Federated Function Serving Fabric for Science},
  booktitle={Proceedings of the 29th International Symposium on High-Performance Parallel and Distributed Computing},
  pages={65--76},
  year={2020},
  doi={10.1145/3369583.3392683}
}

@inproceedings{jette2023slurm,
  author={Jette, Morris A. and Wickberg, Tim},
  title={Architecture of the {Slurm} Workload Manager},
  booktitle={Job Scheduling Strategies for Parallel Processing},
  pages={3--23},
  year={2023},
  publisher={Springer},
  doi={10.1007/978-3-031-43943-8_1}
}

@inproceedings{yang2023skypilot,
  author={Yang, Zongheng and Wu, Zhanghao and Luo, Michael and Chiang, Wei-Lin and Bhardwaj, Romil and Kwon, Woosuk and Zhuang, Siyuan and Luan, Frank Sifei and Mittal, Gautam and Shenker, Scott and Stoica, Ion},
  title={{SkyPilot}: An Intercloud Broker for Sky Computing},
  booktitle={20th USENIX Symposium on Networked Systems Design and Implementation},
  pages={437--455},
  year={2023},
  publisher={USENIX Association}
}

@inproceedings{babuji2019parsl,
  author={Babuji, Yadu and Woodard, Anna and Li, Zhuozhao and Katz, Daniel S. and Clifford, Ben and Kumar, Rohan and Lacinski, Lukasz and Chard, Ryan and Wozniak, Justin M. and Foster, Ian and Wilde, Michael and Chard, Kyle},
  title={Parsl: Pervasive Parallel Programming in {Python}},
  booktitle={Proceedings of the 28th International Symposium on High-Performance Parallel and Distributed Computing},
  pages={25--36},
  year={2019},
  doi={10.1145/3307681.3325400}
}

@article{salim2019balsam,
  author={Salim, Michael A. and Uram, Thomas D. and Childers, J. Taylor and Balaprakash, Prasanna and Vishwanath, Venkatram and Papka, Michael E.},
  title={Balsam: Automated Scheduling and Execution of Dynamic, Data-Intensive {HPC} Workflows},
  journal={arXiv preprint arXiv:1909.08704},
  year={2019},
  doi={10.48550/arXiv.1909.08704}
}

@inproceedings{stubbs2021tapis,
  author={Stubbs, Joe and Cardone, Richard and Packard, Michael and Jamthe, Anagha and Padhy, Suresh and Terry, Stephen and Looney, John and Meiring, Joseph and Black, Michael and Dahan, Maytal and others},
  title={Tapis: An {API} Platform for Reproducible, Distributed Computational Research},
  booktitle={Advances in Information and Communication},
  pages={878--900},
  year={2021},
  publisher={Springer},
  doi={10.1007/978-3-030-73100-7_61}
}

@article{hudak2018ondemand,
  author={Hudak, David and Johnson, Doug and Chalker, Alan and Nicklas, Jeremy and Franz, Eric and Dockendorf, Trey and McMichael, Brian L.},
  title={{Open OnDemand}: A Web-Based Client Portal for {HPC} Centers},
  journal={Journal of Open Source Software},
  volume={3},
  number={25},
  pages={622},
  year={2018},
  doi={10.21105/joss.00622}
}

@inproceedings{mao2016deeprm,
  author={Mao, Hongzi and Alizadeh, Mohammad and Menache, Ishai and Kandula, Srikanth},
  title={Resource Management with Deep Reinforcement Learning},
  booktitle={Proceedings of the 15th ACM Workshop on Hot Topics in Networks},
  pages={50--56},
  year={2016},
  doi={10.1145/3005745.3005750}
}

@article{li2024mrsch,
  author={Li, Boyang and Fan, Yuping and Dearing, Matthew and Lan, Zhiling and Richy, Paul and Allcock, William and Papka, Michael},
  title={{MRSch}: Multi-Resource Scheduling for {HPC}},
  journal={arXiv preprint arXiv:2403.16298},
  year={2024},
  doi={10.48550/arXiv.2403.16298}
}

@article{brown2022wait,
  author={Brown, Nick and Gibb, Gordon and Belikov, Evgenij and Nash, Rupert},
  title={Predicting Batch Queue Job Wait Times for Informed Scheduling of Urgent {HPC} Workloads},
  journal={arXiv preprint arXiv:2204.13543},
  year={2022},
  doi={10.48550/arXiv.2204.13543}
}

@misc{anthropic2026science,
  author={{Anthropic}},
  title={{Claude Science}, an {AI} Workbench for Scientists},
  year={2026},
  howpublished={\url{https://www.anthropic.com/news/claude-science-ai-workbench}},
  note={Accessed: August 5, 2026}
}

@inproceedings{czajkowski1998gram,
  author={Czajkowski, Karl and Foster, Ian and Karonis, Nicholas T. and Kesselman, Carl and Martin, Stuart and Smith, Warren and Tuecke, Steven},
  title={A Resource Management Architecture for Metacomputing Systems},
  booktitle={Job Scheduling Strategies for Parallel Processing},
  series={Lecture Notes in Computer Science},
  volume={1459},
  pages={62--82},
  year={1998},
  publisher={Springer},
  doi={10.1007/BFb0053981}
}

@misc{kubernetes2026resources,
  author={{Kubernetes Team}},
  title={Resource Management for Pods and Containers},
  year={2026},
  howpublished={\url{https://kubernetes.io/docs/concepts/configuration/manage-resources-containers/}},
  note={Accessed: August 5, 2026}
}

@misc{tacc2026updates,
  author={{Texas Advanced Computing Center}},
  title={User Updates: System and Service Maintenance Announcements},
  year={2026},
  howpublished={\url{https://tacc.utexas.edu/news/user-updates/}},
  note={Accessed: August 5, 2026}
}

@misc{nersc2026status,
  author={{National Energy Research Scientific Computing Center}},
  title={Center Status and Perlmutter Maintenance Schedule},
  year={2026},
  howpublished={\url{https://www.nersc.gov/users/status}},
  note={Accessed: August 5, 2026}
}

@misc{nersc2026timeline,
  author={{National Energy Research Scientific Computing Center}},
  title={Perlmutter Timeline},
  year={2026},
  howpublished={\url{https://docs.nersc.gov/systems/perlmutter/timeline/}},
  note={Accessed: August 5, 2026}
}

@article{carion2025sam3,
  author={Carion, Nicolas and Gustafson, Laura and Hu, Yuan-Ting and Debnath, Shoubhik and Hu, Ronghang and Suris, Didac and Ryali, Chaitanya and Alwala, Kalyan Vasudev and Khedr, Haitham and Huang, Andrew and others},
  title={{SAM 3}: Segment Anything with Concepts},
  journal={arXiv preprint arXiv:2511.16719},
  year={2025},
  doi={10.48550/arXiv.2511.16719}
}

@misc{anthropic2026claudecode,
  author={{Anthropic}},
  title={{Claude Code} Documentation: Overview},
  year={2026},
  howpublished={\url{https://code.claude.com/docs/en/overview}},
  note={Accessed: July 16, 2026}
}

@misc{openai2026codex,
  author={{OpenAI}},
  title={{Codex CLI}},
  year={2026},
  howpublished={\url{https://github.com/openai/codex}},
  note={Accessed: July 16, 2026}
}

@misc{anthropic2026opus48,
  author={{Anthropic}},
  title={Introducing {Claude Opus 4.8}},
  year={2026},
  howpublished={\url{https://www.anthropic.com/news/claude-opus-4-8}},
  note={Accessed: July 16, 2026}
}

@misc{openai2026gpt55,
  author={{OpenAI}},
  title={Introducing {GPT-5.5}},
  year={2026},
  howpublished={\url{https://openai.com/index/introducing-gpt-5-5/}},
  note={Accessed: July 16, 2026}
}

@inproceedings{chard2017globus,
  author={Chard, Kyle and Foster, Ian and Tuecke, Steven},
  title={Globus: Research Data Management as Service and Platform},
  booktitle={Proceedings of the Practice and Experience in Advanced Research Computing 2017 on Sustainability, Success and Impact},
  pages={1--5},
  year={2017},
  doi={10.1145/3093338.3093367}
}

@misc{tacc2026policy,
  author={{Texas Advanced Computing Center}},
  title={{Good Conduct on TACC's HPC Systems}},
  year={2026},
  howpublished={\url{https://docs.tacc.utexas.edu/basics/conduct/#ai}},
  note={Accessed: July 06, 2026}
}

\section*{Artifact Description Appendix}

The artifact contains the Diamond Agent implementation, typed agent operations, persistent task/group/event records, and analysis scripts. Sanitized evidence includes 543 federation snapshots, 19 matched-submission rounds with 83 jobs, three AI-generated ResourceEnvelopes with 33 sizing jobs, a live cross-site task-group timeline, a controlled adaptation trace, and two coding-agent traces. These products support the unified-observation, resource-intent, placement, autonomous-adaptation, and agent-continuation claims. Offline analysis needs no scheduler access; live reproduction requires user accounts, allocations, site configuration, and authorized endpoints.

\end{document}